\documentclass[sigconf,authorversion,nonacm]{acmart}

\usepackage{xspace}
\usepackage{url}
\usepackage{subfig}

\usepackage{listings}
\usepackage{caption}
\usepackage{ifthen}
\usepackage[smaller,nolist,nohyperlinks]{acronym}

\usepackage[group-separator=\text{\,},
            group-minimum-digits=4,
            detect-all,
            binary-units=true]{siunitx}

\usepackage{DotArrow}
\usepackage{cancel}

\lstdefinelanguage{LF}{
  keywords={target, deadline, after, state, logical, physical, startup, shutdown, reaction, preamble, x, reactor, trigger, input, output, constructor, new, action, clock, actor, handler, int, main, Main, timer, sec, secs, msec, msecs, usec, usecs, mode, initial, reset, continue,
  behaviortree, sequence, fallback, parallel, task, condition, local, channel},
  keywordstyle=\color{black}\bfseries,
  ndkeywords={class, export, boolean, throw, implements, import, this},
  ndkeywordstyle=\color{darkgray}\bfseries,
  identifierstyle=\color{black},
  sensitive=false,
  comment=[l]{//},
  morecomment=[s]{/*}{*/},
  commentstyle=\color{purple}\ttfamily,
  stringstyle=\color{black}\itshape,
  morestring=[b]',
  morestring=[b]"
}

\newcommand{\lfcode}[1]{\lstinline[language=LF]{#1}}
\newcommand{\code}[1]{\textsf{#1}}

\newcommand{\eg}{e.\,g.\xspace}

\newcommand{\ie}{i.\,e.\xspace}

\newcommand{\etal}{et\,al.\xspace}

\AtBeginDocument{%
  }

\makeatletter
\g@addto@macro\UrlSpecials{\do\!{\linebreak}}
\makeatother

\begin{acronym}
  \acro{api}[API]{Application Programming Interface}
  \acro{agv}[AGV]{Automated Guided Vehicle}
  \acro{bt}[BT]{Behavior Tree}
  \acro{fsm}[FSM]{Finite State Machine}
  \acro{lf}[LF]{Lingua Franca}
  \acro{dsl}[DSL]{Domain Specific Language}
  \acro{ide}[IDE]{integrated development environment}
  \acro{kieler}[KIELER]{KIEL Integrated Environment for Layout Eclipse Rich Client}
  \acro{moc}[MoC]{Model of Computation}
  \acro{uml}[UML]{Unified Modeling Language}
  \acro{scade}[SCADE]{Safety-Critical Application Development Environment}
\end{acronym}

\newcommand{\lfscale}{.55}

\begin{document}

\title[Behavior Trees with Dataflow: Coordinating Reactive Tasks in Lingua Franca]{Behavior Trees with Dataflow:\\Coordinating Reactive Tasks in Lingua Franca}

\author{Alexander Schulz-Rosengarten}
\orcid{0000-0002-1494-8631}
\author{Akash Ahmad}
\orcid{0009-0005-0399-4084}
\author{Malte Clement}
\orcid{0009-0000-8725-6735}
\author{Reinhard von Hanxleden}
\orcid{0000-0001-5691-1215}
\email{{als,stu222517,mac,rvh}!@informatik.uni-kiel.de}
\affiliation{%
  \institution{Kiel University}
  \city{Kiel} 
  \country{Germany}
}

\author{Benjamin Asch}
\orcid{0009-0004-9526-2149}
\author{Marten Lohstroh}
\orcid{0000-0001-8833-4117}
\author{Edward A. Lee}
\orcid{0000-0002-5663-0584}
\email{{benjamintasch,marten,eal}!@berkeley.edu}
\affiliation{%
  \institution{UC Berkeley}
  \city{Berkeley} 
  \country{USA} 
}

\author{Gustavo Quiros Araya}
\orcid{0000-0001-5689-3969}
\author{Ankit Shukla}
\orcid{0000-0002-9434-858X}
\email{{gustavo.quiros,ankit.shukla}!@siemens.com}
\affiliation{%
  \institution{Siemens Technology}
  \country{USA} 
}

\renewcommand{\shortauthors}{A. Schulz-Rosengarten, A. Ahmad, M. Clement, R. von Hanxleden, B. Asch, M. Lohstroh, E. A. Lee, G. Q. Araya, and A. Shukla}

\begin{abstract}
\acp{bt} provide a lean set of control flow elements that are easily composable in a modular tree structure.
They are well established for modeling the high-level behavior of non-player characters in computer games and recently gained popularity in other areas such as industrial automation.

While \acp{bt} nicely express control, data handling aspects so far must be provided separately, \eg in the form of blackboards.
This may hamper reusability and can be a source of nondeterminism.

We here present a dataflow extension to \acp{bt} that explicitly models data relations and communication.
We provide a combined textual/graphical approach in line with modern, productivity-enhancing pragmatics-aware modeling techniques.
We realize and validate that approach in the recently introduced polyglot coordination language \ac{lf}.
\end{abstract}
\acresetall
\acused{api}
\acused{ide}
\acused{uml}
\acused{kieler}

\ccsdesc[500]{Software and its engineering~Model-driven software engineering}
\ccsdesc[300]{Software and its engineering~Abstraction, modeling and modularity}
\ccsdesc[300]{Software and its engineering~Visual languages}
\ccsdesc[500]{Software and its engineering~Orchestration languages}

\keywords{Behavior trees, reactive systems, coordination languages, Lingua Franca}

\maketitle

\section{Introduction}
\label{sec:intro}

\emph{\acp{bt}} originated in the gaming industry, where they are used to program non-player characters~\cite{ColledanchiseO18, MaetasS02}.
They express complex behavior with highly reactive and modular software components coordinating agents in groups.
Their simplicity and modularity has made them increasingly popular in real-world applications as well, such as industrial automation, where \acp{bt} control machines and robots in automated factories.
\acp{bt} use a model-based approach with a simple and intuitive tree structure and a lean set of control flow elements.
\acp{bt} are a high-level composition mechanism that supports the engineering of complex software systems by hierarchically composing atomic behaviors.

We generally believe that \acp{bt} with their largely non-academic heritage
deserve more attention in computer science research than they have received so far. There is
some existing related work
but, overall, the coverage does not seem adequate in light of the practical utility of
\acp{bt} in software engineering.

\subsection{Motivation}
While the simplicity of \acp{bt} is attractive, its minimalist notation leaves the
aspect of \emph{handling data} unaddressed. This aspect, however, is crucial
when using \acp{bt} for implementing behaviors that need to adapt based on
data. In cyber-physical systems, such as robots or of vehicles, data may
originate from sensors that inform the software about the state of its
environment.
A common solution is to use a \emph{blackboard}~\cite{MarzinottoCS14} that
introduces a global set of variables to a \ac{bt}. However, in combination with
a parallel composition in a \ac{bt}, unconstrained access to shared variables
can easily lead to race conditions and non-deterministic behavior. This hampers
reproducibility, robustness, and debugging, and may be fatal when designing
safety-critical software.

Furthermore, as discussed further in \autoref{sec:related-work}, existing \ac{bt} frameworks tend to be graphics-first, where users manually draw \ac{bt} diagrams disconnected from their actual realization, or text-only, where \acp{bt} are realized, \eg, as C++ library.
Both approaches do not fully use the potential of integrating graphics and text.

\subsection{Approach}
To address the handling of data, we propose to adopt a dataflow notation, already exemplified by various modeling languages, such as \acs{scade}~\cite{ColacoPP17} or actors~\cite{AghaMST97}, which breaks down the program into smaller blocks with streams of data flowing between them.
Dataflow, unlike blackboards, explicitly models data interfaces and communication, thus supporting modularity.
Dataflow nodes, or \emph{actors}, have explicit inputs and outputs, which define the way instances of these nodes need to be interconnected with their surrounding context.
\acp{bt} provide a modular design, and instantiating a node or entire \ac{bt} as a child seems rather straightforward, which facilitates reusability.
However, if such a node or \ac{bt} relies on access to a blackboard and may base its behavior on data written by other nodes to these variables, this constitutes a rather brittle interface.
There is typically no indication which variables are considered inputs, outputs, or only local, and how to separate and address multiple instantiations and their memory.
One solution could be to utilize \emph{decorator nodes}, to attach these mappings to the \ac{bt}~\cite{ColledanchiseO18, ColledanchiseN22}.
Instead, we propose to treat communication as a first-class citizen, as modeling it explicitly facilitates formal analyses and program comprehension.

To address the sound integration of text and graphics, we propose a pragmatics-aware modeling approach~\cite{vonHanxledenLF+22}.
Here, all the details and business logic of a model are specified in a textual file, and automatically generated customizable diagrams provide a graphical view to browse and explore the model interactively.

In this paper, we will use \emph{\acf{lf}}~\cite{LohstrohMBL21} as general setting, since we consider \ac{lf} and its open-source infrastructure a good match to address the aforementioned problems.
\ac{lf} is rooted in \emph{reactors}~\cite{LohstrohIRGD+19}, a reactive, event-based, timed-sensitive, and concurrent model of computation with deterministic semantics.
\ac{lf} is a polyglot coordination language in which reactors encapsulate reactive tasks specified in verbatim code.
Reactors provide a high-level \emph{coordination layer} to orchestrate the execution of complex software systems, similar to \acp{bt} coordinating the execution of nodes.
This coordination layer follows a dataflow notation, interconnecting reactor instances.
\ac{lf} supports various platforms and target languages for specifying the low-level business logic, currently including C, C++, Python, TypeScript, and Rust.
The applicability of \ac{lf} ranges from bare metal embedded systems to distributed systems.

The \ac{lf} framework also provides advanced modeling capabilities and tooling for automatically generating customized graphics from a textual file.
E.g., all the \ac{lf} diagrams provided in this paper have been synthesized automatically this way.

\subsection{Contributions and Outline}

\begin{itemize}
  \item We provide
  an industrial use case developed with Siemens Technology
  that poses realistic requirements for a distributed reactive software system, involving the modeling and coordination of autonomous behavior (\autoref{sec:use-case}).
  \item We present a textual and graphical syntax extension of \ac{lf} that seamlessly integrates the \acp{bt} notation into \ac{lf} and employs a pragmatics-aware modeling approach utilizing interactive diagrams (\autoref{sec:syntax}).
  \item We illustrate the structural translation of \acp{bt} nodes into \ac{lf}'s reactors, inspired by an approach by Colledanchise and \"Ogren~\cite{ColledanchiseO18} (\autoref{sec:translation}).
  \item We propose and discuss a notation and semantics for explicitly expressing data relations and deterministic communication in \acp{bt}. This is inspired by and compatible to \ac{lf}'s dataflow and improves the modularity and reusability of data-dependent \acp{bt} (\autoref{sec:communication}).
\end{itemize}
\autoref{sec:discussion} discusses implications of our design and compares it to other approaches in \acp{bt}.
\autoref{sec:related-work} presents further related work.
We conclude in \autoref{sec:conclusions-outlook} and provide a brief outlook on future work.

\section{Use Case: Automated Workplace}
\label{sec:use-case}

To illustrate what we are aiming for with our proposal for \acp{bt} in \ac{lf}, we first present a use case from an industrial context to motivate our solution and introduce \acp{bt} and \ac{lf} in the process.

\subsection{The Scenario}
\label{sec:scenario}

\begin{figure}
  \centering
  \includegraphics[width=.6\linewidth]{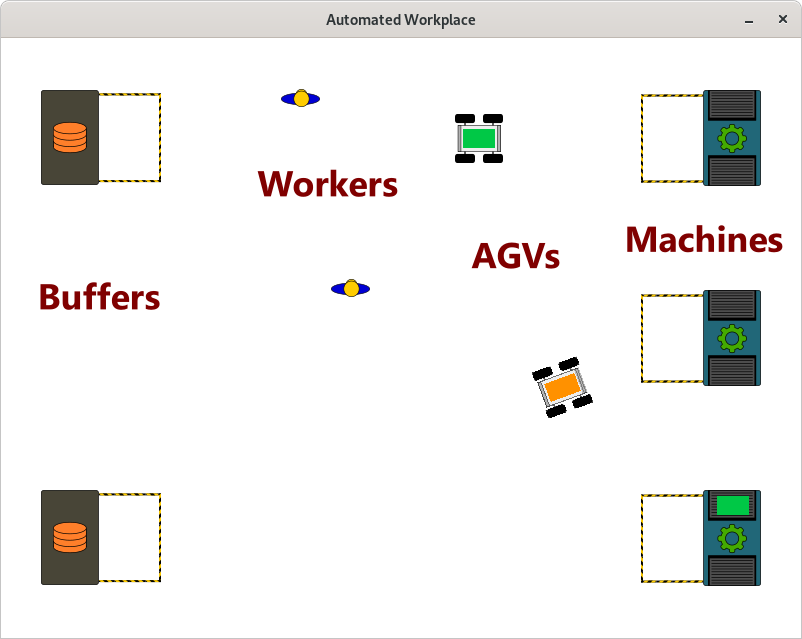}
  \caption{A screenshot of the visualization of a plant simulating the model presented in \autoref{fig:workplace-reactor}}
  \label{fig:workplace}
\end{figure}

We consider an abstraction of a manufacturing plant where \emph{buffers} store resources and intermediate products, \emph{machines} process the goods, and \acp{agv} deliver products between buffers and machines.
Human \emph{workers} perform non-automated tasks and introduce an aspect of safety-criticality to the scenario.
The configuration of number of machines, buffers, and \acp{agv} may vary for different plants, as do the processing times and steps and their composition.
\autoref{fig:workplace} shows a screenshot of our simulation of such a plant with three machines, two buffers, two \acp{agv}, and two workers.

The behavior of each machine is to request delivery of the required resources, process them, which may vary in time based on the type of machine and product, and finally issue the transportation of the produced good to another buffer.
Based on the configuration of the resource requirements of each machine, a sequential manufacturing process can be set up.
A plant may have special input and output buffers which receive resources from outside the plant or ship final products.
For the system itself such goods simply disappear from or emerge in these buffers.
To ensure safety, a machine must stop if a human is in its vicinity, which is detected by sensors on each machine.

The \acp{agv} handle the transportation of goods, they can load and unload products in both buffers and machines.
As mobile units, they must fulfill certain safety properties, namely stopping if in the vicinity of humans, and avoiding crashes with other \acp{agv}.
Their behavior further plays a critical role for the effectiveness of the plant.
Deliveries should be coordinated between \acp{agv}, such that two robots do not fulfill the same task redundantly.
Another aspect is prioritization of machines, which may drastically affect throughput of the entire plant, if machines require different delivery rates.
However, this aspect is not in the focus of this paper.

\smallskip
The task at hand is to create a software system that coordinates the behavior of the entire plant.
As a first attempt, we will model the system as a plain \ac{bt}.
In the following, we use this model for a brief review of \ac{bt} basics.

\subsection{Behavior Tree Basics}
\label{sec:review-btree}

\begin{figure*}
  \centering
  \includegraphics[width=\linewidth]{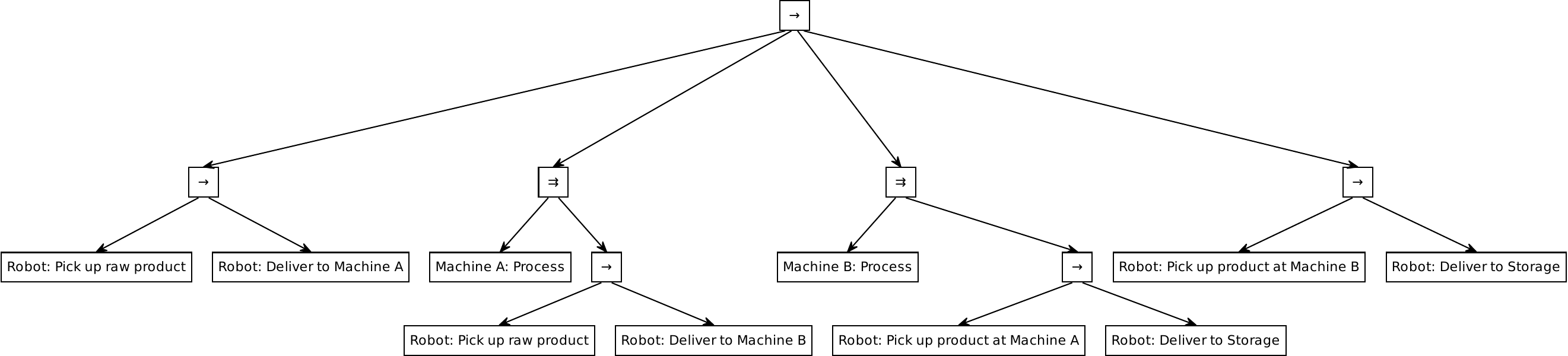}
  \caption{A \ac{bt} that controls an entire plant with a hard-coded processing sequence. 
  }
  \label{fig:workplace-btree}
\end{figure*}
\autoref{fig:workplace-btree} illustrates a \ac{bt} that coordinates the different tasks in a plant setup similar to \autoref{fig:workplace}.
A \ac{bt} consists of \emph{nodes}, usually visualized as labeled rectangles.
Leaf nodes are \emph{tasks}, such as \textsf{Robot: Pick up raw product}, that execute business logic.
After completion, every node returns either \textsc{success}, \textsc{failure}, or \textsc{running} to its parent node.
A \emph{condition} is a special kind of task that only returns \textsc{success} or \textsc{failure} and is rendered as an elliptic node.

\acp{bt} offer three types of compound nodes, \emph{sequence}, \emph{fallback}, and \emph{parallel}, which control the execution of their children.
In a sequence node, if a child returns \textsc{success}, the next child is executed.
If there is no further child to execute, the sequence node returns \textsc{success}.
If a child returns \textsc{running} or \textsc{failure}, the sequence stops there and immediately returns \textsc{running} or \textsc{failure}. 
As illustrated in \autoref{fig:workplace-btree}, sequence nodes are indicated by the single arrow label ($\rightarrow$) and start subsequent tasks upon \textsc{success} of previous ones.
Fallback, labeled with a question mark ($?$), is symmetric by sequentially executing child nodes iff the previous node returns \textsc{failure}. 
In a parallel composition, children are executed concurrently, represented by a node with a double arrow ($\rightrightarrows$).
The return value of the parallel node is based on the child responses and a threshold $M$ specified for the node.
It returns \textsc{success} if $M$ children return \textsc{success}, it returns \textsc{failure}
if $N - M + 1$ children return \textsc{failure}, with $N$ as the number of children, and otherwise it returns \textsc{running}.

\subsection{Discussion of the Plain Behavior Tree Solution}
\label{sec:workplace-btree}

The \ac{bt} in \autoref{fig:workplace-btree} uses sequential composition to define the processing sequence in a plant.
It first orders the delivery of raw resources to \textsf{Machine A} and then supplies \textsf{Machine B} while \textsf{Machine A} is concurrently processing.
Picking up final products is structured analogously.

This \ac{bt} only implements a reduced processing scenario with one robot and two machines.
Furthermore, we simplified this model a bit and left out conditions that check if a tasks in the sequence already finished.
The model shows only one production run for two goods in parallel and assumes that the \ac{bt} is restarted when finished.

While this solution provides the desired overall behavior for our plant example, it hard-codes the overall processing sequence, including \ac{agv} deliveries, in its structure.
This makes it only applicable to this specific plant setup and impedes attempts for re-configuring for different products or scaling up the number of \acp{agv}.

Hence, let us design a system where we model the behavior of machines and robots individually and more generically, independent of the plant configuration.
Given the physical separation of \acp{agv} and machines, this constitutes a distributed embedded system that requires some form of communication between the different devices, \eg over a wireless network.
One could use parallel nodes to model the distributed nature in \acp{bt}, and there are solutions that implement event-driven communication inside \acp{bt}~\cite{AgisGG20}.
However, we propose to utilize a dedicated coordination machinery for this task, one that is specifically tailored to model distribution and communication, in our case \acl{lf}.
Again, we will first review the basics of \ac{lf}, before discussing our running example in detail.

\subsection{Reactor-Oriented Programming}
\label{sec:lf-basics}

\begin{figure}
  \centering
  \includegraphics[width=1\linewidth]{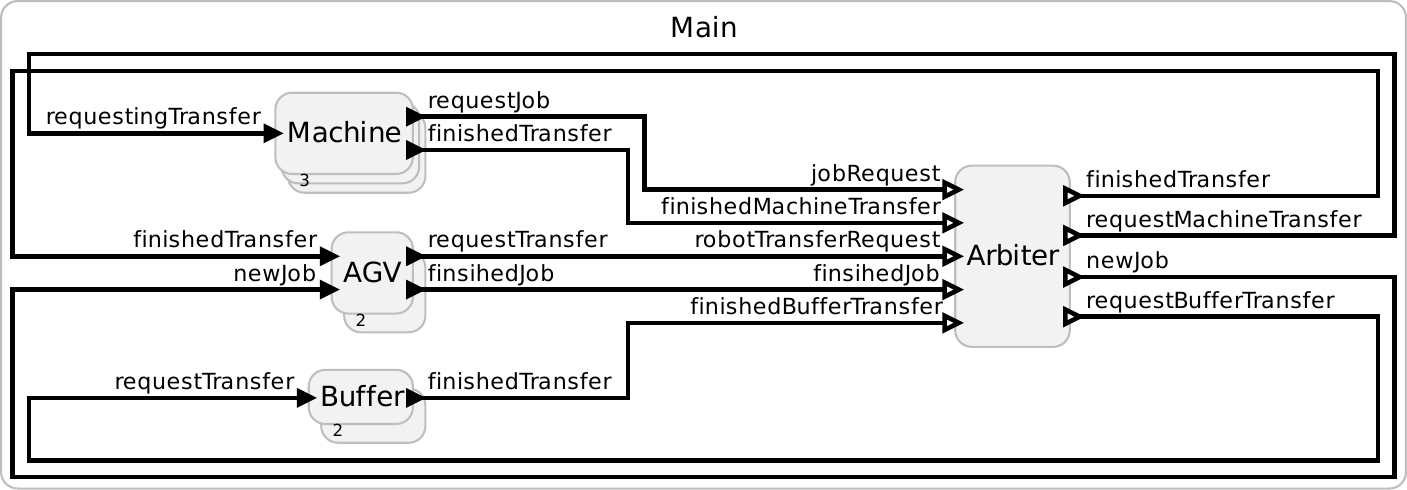}
  \caption{The top-level reactor representing the entire program.}
  \label{fig:workplace-reactor}
\end{figure}

\autoref{fig:workplace-reactor} illustrates the top-level structure of the system as an \ac{lf} program.
The rounded-rectangles represent \emph{reactors}~\cite{LohstrohIRGD+19,LohstrohMBL21}.
\ac{lf} employs the reactor-oriented programming paradigm that is based largely on well-established principles, such as object orientation~\cite{Stroustrup87}, actors~\cite{Hewitt77}, event-driven systems~\cite{DabekZKMM02}, flow-based concepts~\cite{Conway63}.
Reactors declare input and output \emph{ports} that are connected to pass events between them.
These event messages are timestamped and may carry a payload.
The reactor semantics uses a notion of logical time~\cite{LohstrohMSR+20} where event processing happens instantaneously on this logical timeline while physical time progresses normally.
In combination with explicit data dependencies, this allows \ac{lf} to provide deterministic and time-sensitive program execution in single-threaded, multi-threaded, or distributed modes.

In this example, the program contains four interconnected reactors.
An \textsf{Arbiter} acts as a broker that collects requests from machines, assigns jobs to robots, and handles the signaling for loading and unloading.
The \textsf{Machine}, \textsf{AGV}, and \textsf{Buffer} reactors are instantiated as banks~\cite{MenardLBC+23}, meaning that there is an array of multiple instances.
In the diagram they are rendered as stacks with the size of the bank in the left bottom corner.

This model now effectively represents the general structure of the system and the messaging channels between the different components.
Next, we will take a closer look at the \textsf{AGV} reactor to inspect how the behavior of each of the robots is modeled.

\begin{figure*}
  \centering
  \includegraphics[width=\linewidth]{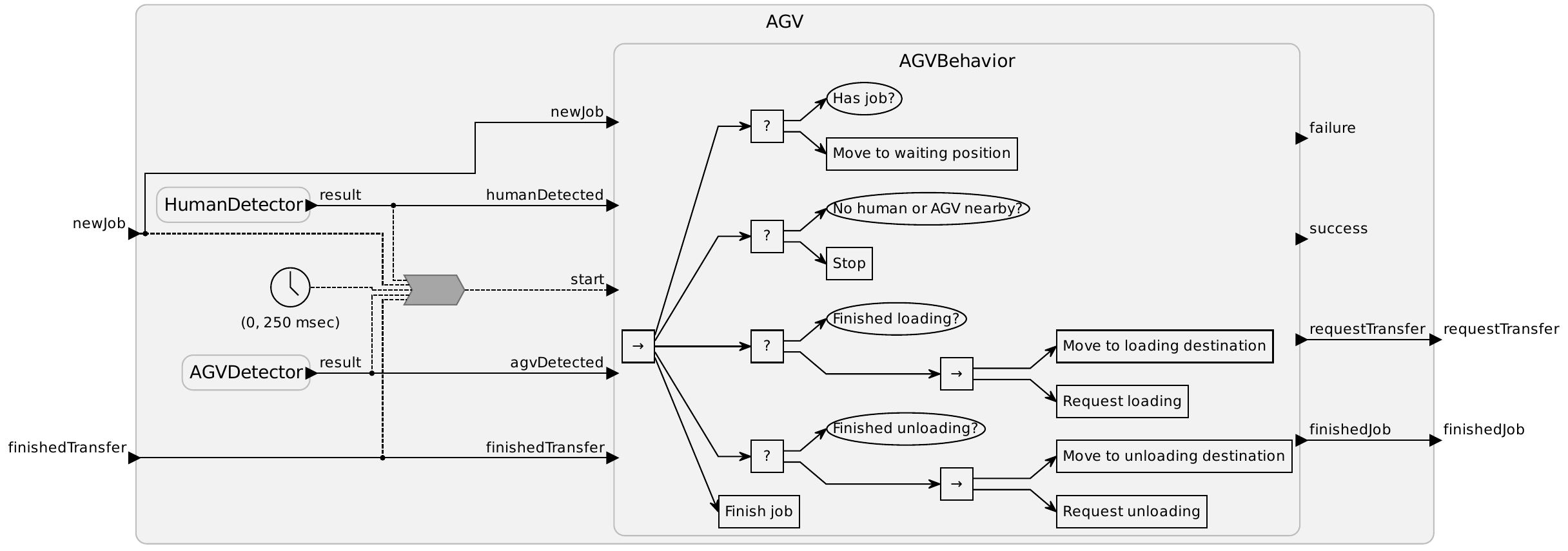}
  \caption{The \textsf{AGV} reactor including the inner \ac{bt} specifying its behavior.}
  \label{fig:agv-reactor-bt}
\end{figure*}

\autoref{fig:agv-reactor-bt} shows the contents of the \textsf{AGV} reactor.
It instantiates three inner reactors, \textsf{HumanDetector}, \textsf{AGVDetector}, and \textsf{AGVBehavior}, as well as a \emph{timer} (the clock figure) and a \emph{reaction} (the gray flag).
Reactions are event-handlers that contain code written in the target language of an \ac{lf} program, \eg C or Python.
They declare triggers, which will invoke their execution if an event occurs, sources to read additional data from, and effects in the form of new events.
This reaction is triggered by the various input ports of the reactor, the results from the two detector reactors, and the timer that will periodically create events every 250 ms with an initial offset of 0.
Its purpose is to issue a reevaluation of the \ac{agv}'s behavior by sending a \textsf{start} signal to the \textsf{AGVBehavior} reactor instance.

\subsection{Discussion of the Reactor-Oriented Solution}
\label{sec:workplace-reactor}

The \textsf{HumanDetector} and \textsf{AGVDetector} reactors have to analyze sensors data and derive a decision whether a human or \ac{agv} is nearby, \eg by utilizing a neural network implemented via a library in the target language.
Thus, a classical dataflow design with ``pure'' \ac{lf} is suitable for them.

The \textsf{AGVBehavior} reactor is more behavior-focused, thus there is a good motivation for expressing it as \ac{bt}.
Note that technically, we can also express \textsf{AGVBehavior} in pure \ac{lf}, as also the translation in \autoref{sec:translation} will illustrate.
However, that derived dataflow-only version is considerably more involved and less abstract than the \ac{bt} variant.

The \ac{bt} rendered in the \textsf{AGVBehavior} 
coordinates tasks similar to reactions in a reactor, but instead of event handling, it uses the execution structure of \acp{bt}.\footnote{Our automatic diagram synthesis allows to switch freely between horizontal and vertical layouts.
In this example, a traditional top-down representation of the \acp{bt} would be rather space consuming.
Thus, we here chose to render the \ac{bt} in left-to-right direction, where child nodes are accordingly ordered top to bottom.}
The root of the tree is a sequence node.
Its first (top-most) child is a fallback node with a condition that checks if the \ac{agv} has currently a job assigned.
This condition also reads the input \textsf{newJob} to accept new jobs from the \textsf{Arbiter}, but this data handling aspect will be discussed in more detail later on.
If the \ac{agv} has no job, the fallback will start the task moving the robot to an waiting position in the plant.
Otherwise, it will return \textsc{success} and the sequences continues by checking if any humans or other \acp{agv} are nearby, with a \textsc{failure} resulting in stopping the \acp{agv}.
The next two subtrees handle picking up and delivering a product and are structured similarly.
First a condition node checks if this part of the job was already finished.
Otherwise the \ac{agv} will move to the (un)loading destination and request a transfer from the machine or buffer.
In the end the \ac{bt} signals the completion of the job to the \textsf{Arbiter}.
Similar to the structure in the \textsf{AGV} reactor, the \textsf{Machine} reactor is also specified using \acp{bt}, but omitted here for brevity.

\smallskip
With this design, \acp{agv} are realized as autonomous agents with their own individual behavior trees. 
This allows to easily scale the number of instances of machines.
Also, with the modular nature of \acp{bt}, more complex systems can be represented using existing behavior trees as building blocks. 
Therefore, \acp{bt} present an elegant solution for modeling such use cases.

\section{Behavior Trees in Lingua Franca}
\label{sec:syntax}

The basic idea of \acp{bt} in \ac{lf} is that a \ac{bt} should be a new kind of reactor, whose inner behavior is coordinated by a \ac{bt} structure instead of a classical reactor composition.
A \ac{bt} reactor should be admissible wherever a normal reactor can be used in an \ac{lf} program.

This integration was already illustrated by the \ac{lf} diagrams in \autoref{fig:workplace-reactor} and \autoref{fig:agv-reactor-bt}.
In \ac{lf} interactive diagrams are provided to the user to perceive and explore the model.
The model itself is specified in a textual source language, as envisioned by the underlying pragmatics-aware modeling approach~\cite{vonHanxledenLF+22}.
Consequently, our integration of \acp{bt} covers both textual and graphical notations.

\subsection{Diagrams}
\label{sec:diagrams}

\ac{lf} uses automatically generated interactive diagrams to illustrate the structural aspects of programs and reactors.
\acp{bt} are integrated into this graphical notation in the form of classical tree structures, as depicted in \autoref{fig:agv-reactor-bt}.
In both \ac{lf} and \acp{bt}, this graphical depiction abstracts from the actual program definition.
For example, \ac{lf}'s reactions (by default) do not show the code that will be executed when triggering such a reaction.
Likewise, the task node of a \ac{bt} only depicts a label that describes the task rather than the actual code.
However, diagrams in \ac{lf} can be interactively configured to show different levels of detail or alternatives variants of certain aspects.
For \ac{bt} reactors, a user can switch to the transformed reactor implementation of a \ac{bt}, described later in \autoref{sec:translation}, or can chose between horizontal or vertical layout direction of \acp{bt}, as illustrated in \autoref{fig:agv-reactor-bt}.

\subsection{Textual Syntax}
\label{sec:textual}

\begin{lstlisting}[
  float,
  language=LF,
  caption={Shortened excerpt of the textual definition for the \textsf{AGVBehavior} \ac{bt} in \autoref{fig:agv-reactor-bt}},
  label=lst:agv-behavior
]
behaviortree AGVBehavior {
  input newJob
  ... further IO ports ...

  sequence {|\label{line:agv-behavior-seq}|
    channel loadingDest|\label{line:agv-behavior-local1}|
    channel unloadingDest|\label{line:agv-behavior-local2}|
    
    fallback {|\label{line:agv-behavior-fb}|
      condition "Has job?" {|\label{line:agv-behavior-cond-state}|
        state currentJob|\label{line:agv-behavior-state}|
        reaction newJob -> loadingDest, unloadingDest {= ... |\textit{code}| ... =}|\label{line:agv-behavior-cond-react}|
      }
      task "Move to waiting position" {= ... |\textit{code}| ... =}|\label{line:agv-behavior-task}|
    }
    fallback {
      condition "No human or AGV nearby?" humanDetected, agvDetected 
      {= ... |\textit{code}| ... =}|\label{line:agv-behavior-cond}|
      task "Stop" {= ... |\textit{code}| ... =}
    }
    ... remaining BT nodes ...
  }
}
\end{lstlisting}

\autoref{lst:agv-behavior} shows an excerpt of the textual source code for the \textsf{AGVBehavior} reactor in \autoref{fig:agv-reactor-bt}.
Instead of a classical reactor, it is declared as \code{behaviortree} that switches the admissible syntax to \acp{bt}.
As a regular reactor, 
it declares input and output ports for communication with other reactors.
We will elaborate on the aspects of communication and data handling in \autoref{sec:communication}.
Additionally, each \ac{bt} reactor has an implicit input port \emph{start} which will trigger the execution of its behavior, as well as \textsc{success}, \textsc{failure}, and \textsc{running} output ports that do not need to be declared explicitly either.
\autoref{fig:agv-reactor-bt} shows these implicit ports, \autoref{sec:translation} will provide more details on their role in the execution.

For control nodes, such as \code{sequence} or \code{fallback}, we use a simple and common nested block notation to encapsulate the child nodes in these control flow structures, see line \ref{line:agv-behavior-seq} and \ref{line:agv-behavior-fb}.
The leaf nodes represent the tasks that are executed.
As such, they correspond to reactions in classical \ac{lf}.
The \ac{bt} notation itself is relatively agnostic to implementation aspects, hence we can reuse many of the syntactic elements and concepts of \ac{lf}'s reactions.
A task node, as in line \ref{line:agv-behavior-task}, is declared via
\lfcode{task "<label>" <sources*> -> <effects*> \{=<code>=\}}.
While in classical reactions a label is optional and defined with a slightly different syntax, the main difference is that tasks do not declare a list of triggers that would issue the execution of a reaction.
Instead, tasks will be executed based on the semantics of the enclosing control node, see \autoref{sec:translation}.
The list of none to many data sources this task may read from and the optional list of effects corresponds directly to the \emph{causality interface} of reactions~\cite{LohstrohIRGD+19}.

\ac{lf} uses a language-independent way of integrating with its target language that separates the coordination layer from the implementation.
It allows the user to write target code directly inside the \code{\{=...=\}} brackets but will not further analyze this code and handle it under a black-box abstraction.
In turn, this enables \emph{polyglot} designs~\cite{LohstrohMBL21}.
To ensure determinism in the \ac{lf} execution and to provide the code with access to relevant data communicated on the reactor level, the reactions have causality interfaces that declare the access to ports and other data sources, as well as define potential effects the code may emit during execution.
Our task syntax uses the same handling of target code in combination with causality interfaces.
In fact, as \autoref{sec:translation} will describe, tasks will be transformed into classical reactions, which benefits from this correspondence.

We further support the definition of a \code{condition}, which is essentially a task but not expected to yield a \textsc{running} response.
Primarily, this affects the graphical representation, switching to a elliptic rather than rectangular node shape, but otherwise conditions are handled like regular tasks.
The conditions in lines \ref{line:agv-behavior-cond-state} and \ref{line:agv-behavior-cond} illustrate a few more variations of the available syntax.
Line \ref{line:agv-behavior-cond} is similar to the task in line \ref{line:agv-behavior-task} but declare \textsf{humanDetected} as data source that may be accessed at execution.
The condition in \ref{line:agv-behavior-cond-state} uses an extended syntax for tasks and conditions that allows defining additional elements.
These will go into the reactor that later represents the node, see \autoref{sec:translation}.
Instead of directly defining the causality interface and target code for the conditions, a new code block is opened in which a state variable, here \textsf{currentJob}, is defined.
The \textsf{reaction} keyword then starts the definition of the business logic for the condition.
As the causality interface reflects, it will check whether the \textsf{Arbiter} assigns the \ac{agv} a new job, given that it is idle at the moment, and it will have two effects that specify the target for picking up the cargo and the destination for delivery.
These two effects are special communications channels inside the \ac{bt} that we introduce alongside our proposed concept.
\autoref{sec:locals} will describe their functionality in more detail, but in this example it allows the \textsf{Has Job?} condition to supply the tasks that later move the \ac{agv} with the job-specific destinations.
Its stateful nature allows to store the job, while the communication to the \textsf{Arbiter} only happens upon changes, as an event-driven nature would suggest.

While in this case the condition only introduces a single state variable that its reactor will carry, our proposed syntax permits to define any additional reactor contents, including instantiating reactors.
This approach results in a hybrid design combining the \ac{bt} notation and \ac{lf}.

\subsection{Target Language Interface}
\label{sec:interface}

In accordance to the design of \ac{lf}, our task nodes allow the implementation of a task's behavior in a desired target language, directly inside the task declaration.
Of course, the programmer can also call externally defined functions and does not need to implement everything directly in an \ac{lf} file.
\ac{lf} automatically generates a target language interface that provides access to sources and effects, declared in the interface of a reaction, or in this case tasks.
For \ac{bt} tasks, however, there are two effects that do not need to be specified explicitly: \textsc{success} and \textsc{failure}.
These correspond to the general interface of \ac{bt} nodes and are implicitly present.
A task implementation is expected to produce these effects to influence the way the reaction of a \ac{bt} is executed.
In our implementation for \acp{bt} in \ac{lf}, we use this simplified interface, omitting the \textsc{running} response.
We consider the absence of \textsc{success} and \textsc{failure} as \textsc{running}.
This is enabled by the deterministic semantics of \ac{lf}, which ensures a clearly defined state for all inputs before computing any results based on these inputs.
This simplification also eases the transformation described in \autoref{sec:translation}.
Nonetheless, our concept does not rely on this design and could easily be extended to explicitly use \textsc{running}.

The polyglot approach of \ac{lf} makes our proposed concept for \acp{bt} in \ac{lf} available to a variety of target languages, currently C, C++, Python, Typescript, and Rust.
However, the aforementioned black-box perspective of target code makes it the responsibility of the programmer to adhere to the correct usage of the \textsc{success}-\textsc{failure}-\textsc{running}-protocol.
For example, the code must not produce both the \textsc{success} and \textsc{failure} output in the same tick.
We will revisit this issue as room for future work in~\autoref{sec:conclusions-outlook}.

\section{Structural Translation}
\label{sec:translation}

\begin{figure}
  \centering
  \includegraphics[width=.6\linewidth]{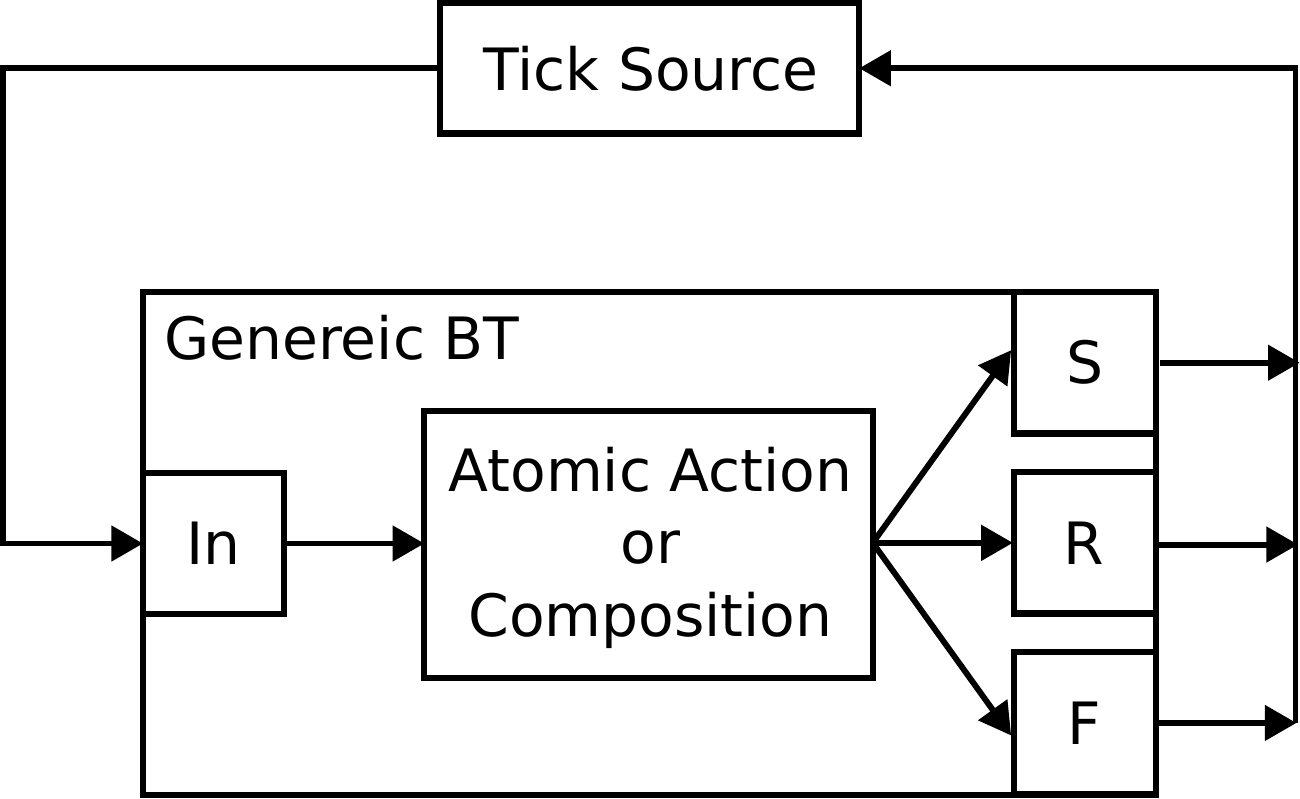}
  \caption{General structure of a \acs{fsm} translation of a \ac{bt}, adapted from Colledanchise and \"Ogren~\cite{ColledanchiseO18}}
  \label{fig:fsm-bt}
\end{figure}

So far we only illustrated the syntactic elements of our \ac{bt} extension to \ac{lf}.
Now we take a look at how to implement the new constructs.
We could utilize existing libraries for \acp{bt}, \eg Py Trees, 
and use the target language integration of \ac{lf} to translate a \ac{bt} into executable code.
However, then the entire \ac{bt} would be represented by a single reaction that in case of using parallel nodes could not benefit from \ac{lf}'s existing capabilities for parallel execution.
We also considered different pattern-based schemata that would utilize \ac{lf}'s modes or represent \acp{bt} by various reactions in a reactor.
However, we here propose a concept inspired by the ``\ac{fsm} translation'' for \acp{bt} by Colledanchise and \"Ogren~\cite{ColledanchiseO18}.

\autoref{fig:fsm-bt} illustrates their general idea.
A \textsf{Tick Source} starts the execution of a reaction.
Then the system transitions to the \textsf{In} state of the root behavior tree node.
Afterwards it switches to the inner behavior, which is either the execution of an action (is case of a task node) or transitions to an \textsf{In} state of an inner node.
If the inner composition is a sequence, the \textsc{success} state of the first child will lead to the \textsf{In} state of the second child node modeling the semantics of executing children sequentially.
After transitioning through the inner nodes the system will end up in either the \textsf{S}, \textsf{R}, or \textsf{F} state, reflecting the return value of the node.

As a first observation, while this node/edge diagram syntactically resembles an \ac{fsm} representation, we argue that it actually does not represent an \ac{fsm}, at least not at the abstraction level of \acp{bt}.
Even if we would consider the nodes as ``states'' (rather than actors), these would represent the steps of the execution \emph{within} a tick, rather than the states of the system that persist across ticks.
Thus, we do not follow the argument presented in \cite{ColledanchiseO18} that this construction shows, roughly, the conceptual equivalence of \acp{bt} and \acp{fsm}.

However, putting aside the original \ac{fsm} interpretation of this construction, it quite closely resembles a dataflow composition, which opens the door to the mapping we propose here.
If a token travels across the pathways of the ``transitions'' (connections) and triggers reactions that execute tasks, it corresponds well to \ac{lf}.
This constitutes the basic principle for our structural translation approach.

\begin{figure}
  \centering

  \subfloat[General structure and task/condition]{
    \centering
    \includegraphics[scale=\lfscale]{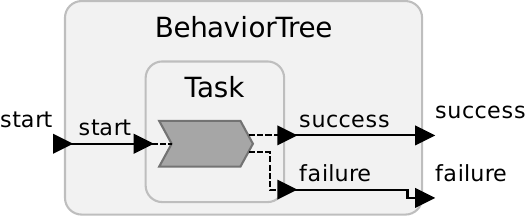}
    \label{fig:translation-task}
  }

  \subfloat[Sequence]{
    \centering
    \includegraphics[scale=\lfscale]{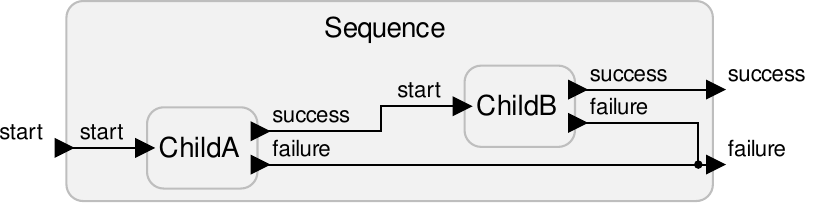}
    \label{fig:translation-sequence}
  }

  \subfloat[Fallback]{
    \centering
    \includegraphics[scale=\lfscale]{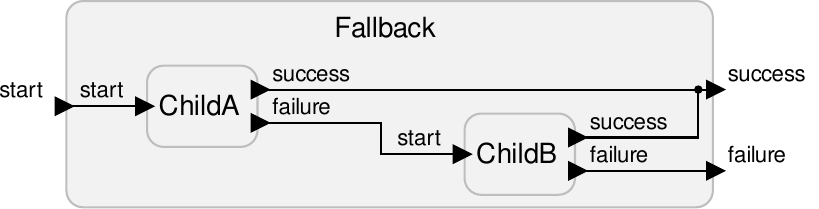}
    \label{fig:translation-fallback}
  }

  \subfloat[Parallel]{
    \centering
    \includegraphics[scale=\lfscale]{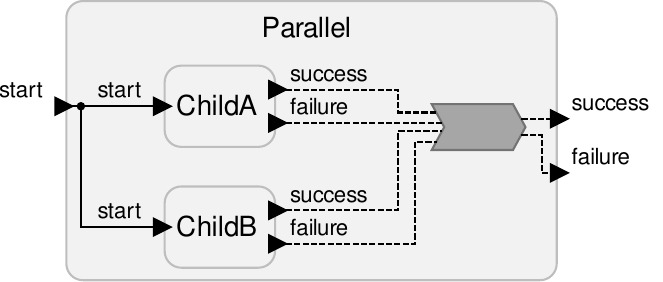}
    \label{fig:translation-parallel}
  }

  \caption{Patterns of the \ac{bt} translations into reactors.}
  \label{fig:translation-nodes}
\end{figure}

We translate each node into a reactor individually.
Hierarchical nesting and connections then result in a structure that will coordinate the tasks in the correct way.
\autoref{fig:translation-nodes} provides an overview on the resulting reactors for each node type.

\subsection{General Structure}

\autoref{fig:translation-task} depicts the reactor for the entire \ac{bt} to illustrate the general structure, in addition to the task translation.
Each reactor resulting from a node has a start input port and \textsf{success} and \textsf{failure} output ports.
As illustrated here, we also expose them on the top level.
Hence, compared to \autoref{fig:fsm-bt}, we locate the \textsf{Tick Source} outside the \ac{bt} reactor in the event-driven domain of \ac{lf}.
\autoref{fig:agv-reactor-bt} is an example where the \ac{bt} reaction is, among others, triggered by another reactor for detecting humans via hardware sensors.
The further execution of the \ac{bt} is coordinated by forwarding this start event correctly.
Note that in the \ac{lf} semantics, this happens (in the absence of explicit delays) instantaneously, \ie without advancing logical time. 

\subsection{Task \& Condition}

\autoref{fig:translation-task} shows the translation of a single task.
The body of the task is converted into a reaction that is triggered by \textsf{start}.
This will execute the embedded code and produce a new output at either the \textsf{success} or \textsf{failure} port of the task's reactor.
If no such output is produced, we consider the output \textsc{running}.
The effect of a running task is that no subsequent tasks should be executed, except in a parallel execution, see below.
This perfectly matches the situation where none of the ports can trigger further behavior.

Condition nodes are handled exactly the same.

\subsection{Sequence \& Fallback}

A sequence composes two or more nodes, which can be task nodes or composite nodes representing an entire subtree.
\autoref{fig:translation-sequence} shows the reactor for a sequence of two nodes of any kind.
The \textsf{start} port of the \textsf{Sequence} reactor will immediately start the \textsf{NodeA}.
However, \textsf{NodeB} will only start if \textsf{NodeA} returns \textsc{success}, implemented by connecting the \textsf{success} output of \textsf{NodeB} to the \textsf{start} of \textsf{NodeB}.
The \textsf{Sequence} reactor itself will only return \textsc{success} based on \textsf{NodeB} but will yield a \textsc{failure} as soon as any of the nodes indicated this result.
Again, the absence of \textsc{success} or \textsc{failure}, \ie \textsc{running}, will stop the triggering of downstream reactors.

\autoref{fig:translation-fallback} uses a similar structure, but starts subsequent reactors when previous ones return \textsc{failure}.
The dataflow notation of \ac{lf} very clearly illustrates the semantics of \acp{bt} in this case.

\subsection{Parallel}

In a parallel composition, as depicted in \autoref{fig:translation-parallel}, the start event is broadcast to all children.
Their \textsc{success} and \textsc{failure} responses, however, are jointly handled by a special  reaction.
This response of the parallel node is subject to some logic that is generated by the compiler during translation.
A parallel node specifies a threshold $M$ that defines how many of its children must return \textsc{success} in order to yield \textsc{success} itself.
The code in the reaction collects the responses from the child nodes, checks whether the threshold $M$ is met or not, and produces the corresponding output.

As the diagram illustrates, \ac{lf} reactors can quite naturally express a parallel composition.
If the program is compiled with the multi-threaded runtime support enabled, \ac{lf} will also parallelize the execution of these reactors.
While this concurrent compositions is a common feature in dataflow languages, the design by Colledanchise and \"Ogren did not include a translation for parallel nodes because the classical model of \acp{fsm} cannot express this behavior.

\section{Communication}
\label{sec:communication}

\autoref{sec:translation} illustrates the general patterns of our translation approach.
However, it does not show the connections that are necessary to convey data between reactors, as these are application specific.
Moreover, as discussed, the aspect of communication and data handling is in general not included in the \ac{bt} notation at all, and usually handled via additional concepts, such as a blackboard~\cite{MarzinottoCS14}.
In contrast to that, this aspect is central to a dataflow notation.
We argue that with a combination of both notations, as proposed here, we can improve the way data is handled in \acp{bt} and indirectly also improve their modularity and reusability.
At the same time dataflow-oriented languages, such as \ac{lf}, are enhanced by a more compact and effective control flow notation.

\subsection{Input \& Output}
\label{sec:io}

Each reactor in \ac{lf} has a clear input output interface, manifested in its ports.
In the same spirit, we add ports to the \ac{bt} syntax in \ac{lf}, as discussed in \autoref{sec:textual}, and allow tasks to specify additional sources and effects that will access and communicate through these ports.

On the language design side, we do not require the user to repeat this interface at each level of hierarchy (node nesting), as our example in \autoref{lst:agv-behavior} shows, but allow tasks to directly access the port defined on the behavior tree.
Yet, the reactors we create during translation require port declarations for each of them.
Hence, this task is automated in the compiler.
If nested tasks require access to some input or output port, each level of reactor will have the necessary ports and connections to forward the data.

\subsection{Internal Communication via Channels}
\label{sec:locals}

In addition to communication with the environment, it is also common to share data between nodes of the same \ac{bt}.
As already mentioned, globally shared variables, \eg via a blackboard, can be a source of nondeterminism.
Moreover, they can decrease modularity if data-dependencies are only modeled implicitly and moving nodes or factoring out a subtree leads to unnoticed breaks in data relations.

Hence, we propose to use communication \emph{channels}, which is an established dataflow concept, to address this issue.
In \ac{lf}, we realize channels with ports and connections.
Similar to the approach with inputs and outputs in \acp{bt} (\autoref{sec:io}), we want to relieve the user from the burden of specifying these communication interfaces at each level of hierarchy, as it is required for reactors.
Instead, we introduce a \textsf{channel} syntax for control nodes in our \acp{bt}, see lines \ref{line:agv-behavior-local1} and \ref{line:agv-behavior-local2} in \autoref{lst:agv-behavior}.
This can be referenced by any task or condition inside this node as either source or effect.
The transformation will automatically create the necessary input or output port, depending on the access type in the causality interface, and establishes the connections between readers and writers, also across hierarchies.
With these channels, users can now send messages between nodes in a \ac{bt}, while \ac{lf}'s semantics will reject any nondeterministic outcome in parallel composition.
Additionally, the correspondence to input output ports makes it easy to convert a channel into one of these, if one decided to factor out a subtree into a reusable standalone \ac{bt}.
Furthermore, the efficient implementation of \ac{lf} will usually detect cases where reactors communicate in the absence of concurrency or networks and optimize this case such that there is no real overhead for instantaneously passing messages.
\autoref{sec:dataflow-comparison} discusses sending messages for shared variables.

\begin{figure*}
  \centering
  \subfloat[Textual \ac{bt} syntax]{
    \begin{minipage}[b]{.3\linewidth}
    \centering
\begin{lstlisting}[language=LF]
target Python
behaviortree Forward {
  sequence {
    channel x
    task Writer -> x {= x.set(42)
      success.set_present() =}
    task Reader x {= if x.is_present: print(x.value)
      success.set_present() =}
  }
}
\end{lstlisting}
    \label{fig:local-forward-code}
    \end{minipage}
  }
  \subfloat[Transformed reactors]{
    \begin{minipage}[b]{.7\linewidth}
    \centering
    \includegraphics[scale=\lfscale]{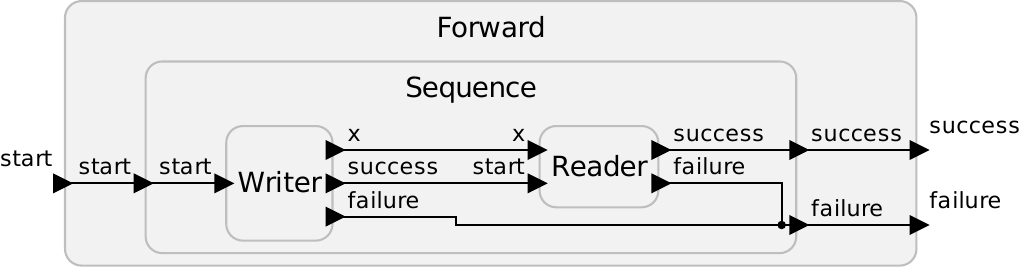}
    \label{fig:local-forward-reactors}
    \end{minipage}
  }

  \subfloat[Textual \ac{bt} syntax]{
    \begin{minipage}[b]{.3\linewidth}
    \centering
\begin{lstlisting}[language=LF]
target Python
behaviortree Backward {
  sequence {
    channel x
    task Reader x {= if x.is_present: print(x.value)
      success.set_present() =}
    task Writer -> x {= x.set(42)
      success.set_present() =}
  }
}
\end{lstlisting}
    \label{fig:local-backward-code}
    \end{minipage}
  }
  \subfloat[Transformed reactors]{
    \begin{minipage}[b]{.7\linewidth}
    \centering
    \includegraphics[scale=\lfscale]{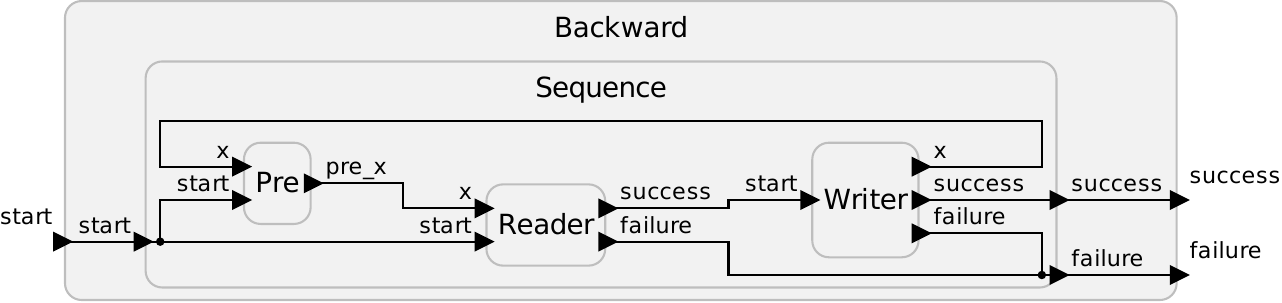}
    \label{fig:local-backward-reactors}
    \end{minipage}
  }
  \caption{Examples of forward and backward communication patterns.}
  \label{fig:local-forward-backward}
\end{figure*}

Regarding the implementation, there are two main cases when creating communication channels inside a \ac{bt}: \emph{forward} and \emph{backward} communication.

\paragraph{Forward Communication}
The simple case for communication is that a task reads the value on a channel after it was written by a previous task, \eg in a sequence.
\autoref{fig:local-forward-code} illustrates this pattern.
In the source code, the first task declares the channel \textsf{x} as an effect and sets its value, while the \textsf{Reader} task reads it.
In the resulting reactors, shown in \autoref{fig:local-forward-reactors}, the \textsf{Writer} reactor has \textsf{x} as an output port directly connected to \textsf{Reader} where \textsf{x} is an input.
Since these two tasks will be executed sequentially at the same tick of the \ac{bt}, the value is passed instantaneously (with the same tag) between the two reactors.
In case of multiple preceding writers, a reader will only receive the latest value written to this channel.
Such a sequential overriding is always possible, while parallel writers will be generally rejected by the \ac{lf} compiler.

\paragraph{Backward Communication}
Cases of backward communication occur when a reader is located sequentially before a writing task, as illustrated in \autoref{fig:local-backward-code}.
One could argue that such a reader cannot receive a message sent by a subsequent task, but in a repetitive reactive setting, it makes sense to allow influencing tasks in a future reaction.
Hence, we also establish backward connections from writers and preceding readers.
However, during execution of such a \ac{bt}, the preceding task was potentially already executed, and immediately executing this tasks again and out-of-order does not make much sense.
Additionally, \ac{lf} does not support executing reactions multiple times at the same tag either and would consider such a direct backward connection a causality cycle.
Therefore, we delay these messages that are sent upstream in execution order by one tick.
This is a common concept, also present in synchronous languages, where the user has access to values of the previous tick.

Since the execution ticks of a \ac{bt} depend in our case on the \textsf{start} input and not some fixed timing delay, as otherwise common in \ac{lf}, we introduce an automatically generated \textsf{Pre} reactor, as shown in \autoref{fig:local-backward-reactors}, to store the value that was written to \textsf{x} in the current tick and reintroduce it with the next triggering via \textsf{start} to reach the reader preceding the writer.
Again, multiple writers are subject to sequential overriding, and instantaneous sequential writers (forward communication) will supersede receiving a previous value.

\subsection{Communication in the AGV Behavior}

\begin{figure*}
  \centering
  \includegraphics[width=\linewidth]{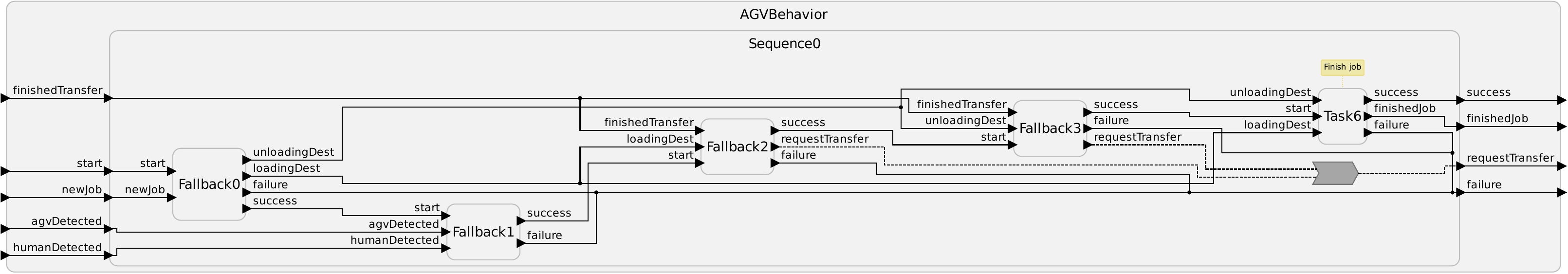}
  \caption{The \ac{bt} of \textsf{AGVBehavior} (\autoref{fig:agv-reactor-bt}) after its translations into reactors. The diagram only has the top-level sequence expanded, while sub tress, such as the fallback nodes, are still collapsed. While this is not the recommended default to illustrate a \ac{bt}, it exposes both the actual implementation and the data connections between the nodes, representing a dataflow view for the \ac{bt}.}
  \label{fig:agv-data}
\end{figure*}

While the \ac{bt} in \textsf{AGVBehavior} does not reflect all communication patterns, inspecting it in its transformed form in \autoref{fig:agv-data} reveals a more detailed picture on how the data is flowing through the nodes of the original \ac{bt}.
The diagram only shows the reactor for the top-level sequence in its expanded form, while subtrees, such as \textsf{Fallback1}, are still collapsed and do not expose their inner elements.
The inputs and outputs cross the hierarchy level of the sequence to reach their tasks or are further forwarded into subtrees representations.
Likewise, the channel connections are established from the \textsf{Has Job?} task inside \textsf{Fallback0} to the two respective subtrees that move to the load and unload destination.

While this illustration only implicitly represents the tree structure of the \ac{bt}, it exposed both the actual implementation and the data connections between the nodes, representing an alternative \emph{dataflow view} for the \ac{bt}.

\section{Discussion}
\label{sec:discussion}

\subsection{Further Behavior Tree Node Types}
\label{sec:advanced}

In our proposed concept, we focused on the basic nodes of \acp{bt}: sequence, fallback and parallel. 
However, the \ac{bt} notation also includes sequence and fallback variants with \emph{memory}.
These nodes behave like their non-memory counterparts but save the execution progress of their children and continue their executions sequence in the next tick.
More specifically, they save the results returned by their children and do not re-tick the subtrees in subsequent ticks until the node evaluates to \textsc{success} or \textsc{failure} and resets the memory. 
Hence, memory nodes are syntactic sugar for non-memory nodes, with additional conditional nodes that track progress and skip children that have already returned \textsc{success} for sequence, or \textsc{failure} for fallback, respectively.

While we, for now, skipped these two node type in our proposal, we expect that their addition to our proposed transformation into reactors would be rather straightforward. 
Specifically, one might combine state variables in reactors with automatically generated reactions that track the execution progress of a memory sequence or fallback, to control which of the child nodes, or resulting reactors, should be started in a tick.
Alternatively, reactors also feature a mode notation~\cite{SchulzRosengartenvHLB+23} that could be used to model the execution sequence as a state machine.

Another type of nodes that is commonly used in \acp{bt} are \emph{decorators}. 
These customizable nodes can be used to manipulate the return values and execution of other nodes.
The \ac{bt} notation does not prescribe specific decorators and refrains from restricting their functionality in any way.
Nonetheless, there are a few commonly used decorators. 
For example, \textsf{invert} that flips the \textsc{success} and \textsc{failure} response of a node, \textsf{max-N-tries} that lets a child fail a fixed number of times
until it returns \textsc{failure} without ticking the subtree, and \textsf{repeat-N-times}, which executed a subtree multiple times.
While we did not include decorators in our current realization for \acp{bt}, \ac{lf} offers the capabilities to implement such decorators.
For example, \textsf{invert} could be added by inserting a single reaction that flips the return messages, while \textsf{max-N-tries} and \textsf{repeat-N-times} would require stateful reactors for counting execution runs.

\subsection{Data Handling in Classical Behavior Trees}
\label{sec:dataflow-comparison}

As previously described, we realize communication between tasks as channels rather than shared variables, which are commonly used via blackboards.
This slightly changes the way a programmer has to handle data in a \ac{bt}.
While blackboards introduce a set of variables that are implicitly shared between all nodes of a \ac{bt}, we require the explicit declaration of data access in the causality interface of tasks.
With a clear input output interface for each task, we are able to ensure determinism in concurrency through the semantics of \ac{lf}.

Another core difference is that we shift the notion of a state, reflected by variables that keep their values across ticks, from the level of the \ac{bt} to the nodes itself.
Our channels then fill the gap and enable sparse event-based communication between nodes to share data during a tick (or partially reaching into the next tick, in case of backwards communication).
Admittedly, this might appear as an inconvenient overhead to the programmer, especially in small use-cases.
Yet, most software eventually grows large and at least then requires the established principles of software engineering to keep a system maintainable.
In this aspect we argue that our approach improves over blackboards because we fundamentally embrace modularity, which is also a core idea of \acp{bt} \cite{BiggarZS20, ColledanchiseO18}.
The explicit use of interfaces and the node-local notion of state facilitates testability of individual nodes or subtrees, and likewise enables factoring out these subtrees as reusable components.
In particular, the existence of clear interfaces plays a key role when instantiating \acp{bt} inside other \acp{bt}, since it allows a well-defined interaction between the instantiating and instantiated \ac{bt}.
Blackboards would require some form of mapping in order to establish a connection and prevent name clashes in the presence of multiple instances of the same \ac{bt}.
Yet, the role of each variable would remain informal.

Similar considerations for modularity, testability, and reusability also influenced the design of reactors~\cite{LohstrohIRGD+19} and resulted in the current \ac{lf} language that has state variables completely local to reactors not shared with inner reactors, and ports for event-driven communication.
With our approach of implementing \ac{bt} nodes as reactors, we subjected our concept to the restrictions of \ac{lf}.
Nonetheless, we also investigated introducing a shared state concept to \ac{lf}, which would enable a syntax closer to blackboards.
Yet, the results showed similar problems as blackboards.
Hence, we consider our proposed solution to be an adequate compromise for a notation that combines the core benefits of dataflow-oriented designs with \acp{bt}.

\smallskip
A potential drawback of our approach is found in the overhead for communicating data via channels.
While simple cases of one-to-one communication are usually optimized by the \ac{lf} compiler to efficiently access data, \eg via a pointer, more complex channel setups require the introduction of additional reactions, as described in \autoref{sec:locals}.
The reason is that in \ac{lf}, the default case is a concurrent composition of reactors.
Hence, for sequence and fallback, these additional reactions have to transfer the sequential notion of the node's semantics to the event system of \ac{lf}.
While this aspect also makes the graphical view of the generated reactor more complex, our pragmatics-aware modeling approach still allows to switch to the more lean \ac{bt} view to effectively communicate and expose the different aspects in the model to the user.
We also plan to further investigate refinements of these views, for example, by optionally filtering out automatically generated aspects in the reactor diagram or introducing data-flow aspects to the \ac{bt}.

\section{Related Work}
\label{sec:related-work}
The basic notation and first implementations of \acp{bt} are presented in the context of non-player characters for video games~\cite{MaetasS02}.
\acp{bt} have since been adapted for a variety of tasks over multiple domains~\cite{ColledanchiseO18}.
Biggar \etal~\cite{BiggarZS21} developed a formal framework to compare the expressive power of \acp{bt}, Finite State Machines, Teleo-reactive Programs~\cite{Nilsson93}, and Decision Trees~\cite{ColledanchiseO18}.
They also proposed a formalization of the notions of reactiveness and modularity and introduced $k$-\acp{bt} as a natural generalization of \acp{bt}~\cite{BiggarZS20}.
These $k$-behavior trees have $k$ different return values instead of \textsc{success} and \textsc{failure}, which benefits intuitive failure handling.

As discussed, some \ac{bt} realizations use blackboards to store shared data and to create stateful behavior~\cite{ColledanchiseO18,IovinoSSO+22}. 
Agis \etal~\cite{AgisGG20} use an event-driven \ac{bt} semantics which combines blackboards with listeners and task priorities, such that tasks can be aborted or re-evaluated by other nodes.
In their setting, multiple autonomous robots have to communicate with
each other to organize into groups and solve a task together.
In our approach, the event-driven communication inside a \ac{bt} has a more explicit representation in the syntax, and
a well-formed semantical foundation in dataflow.

There are also approaches to mitigate or circumvent the drawbacks of blackboards.
Shoulson \etal proposed an extension for behavior trees with parametrized interfaces~\cite{ShoulsonGJM+11}. 
We naturally embed these parametrized interfaces in the form of input output interfaces. 
Colledanchise and Natale investigated the potential problems with concurrency caused by parallel 
nodes~\cite{ColledanchiseN18, ColledanchiseN22} and introduced concurrent \acp{bt}.
Concurrent \acp{bt} use special decorators such as the \textsf{ProgressSynchronizationDecorator} 
and the \textsf{RessourceSynchronizationDecorator} to synchronize concurrent tasks.

There are established programming models that avoid race conditions on concurrently accessed data, for example in Rust~\cite{MatsakisK14} or synchronous languages~\cite{BenvenisteCEH+03}.
The synchronous model of computation appears quite compatible with the execution semantics of reactive \acp{bt}~\cite{vonHanxledenSRA+22}.
Several languages already combine dataflow aspects with more control-flow-oriented states machines or modes, such \acs{scade}~\cite{ColacoHP06}, SCCharts~\cite{GrimmSSRvH+20}, or as already mentioned \ac{lf}~\cite{SchulzRosengartenvHLB+23}.

In the field of requirements engineering there
is also a \ac{bt} notation used in the Genetic Software Engineering
approach~\cite{Dromey03}, where the goal is to construct a design out of a set
of functional requirements by integrating behavior trees for individual
functional requirements, one-at-a-time, into an evolving design behavior tree.
Despite the same name, the notation is not directly related to the
coordination/programming notation~\cite{ColledanchiseO18} discussed in this
paper.

There is work by Ghzouli \etal also using a robotic use case to evaluate the capabilities of \acp{bt} in terms of safety~\cite{GhzouliBJD+20}.

Concerning the textual/graphical aspects of \acp{bt},
game engines typically represent \acp{bt} graphically, \eg Unity~\cite{Barrera18} and Unreal Engine\footnote{\url{https://docs.unrealengine.com/5.0/en-US/behavior-trees-in-unreal-engine/}}.
This usually works with a palette-based workflow, where nodes are partially predefined by the tool.
However, embedded devices often require a more low-level implementation with specific hardware libraries.
There are also \ac{bt} libraries, such as Py Trees\footnote{\url{https://py-trees.readthedocs.io}}, that provide an \ac{api} to programmatically compose \acp{bt} and execute them.
Yet, these usually lack the support for graphical visualization.

\section{Conclusion}
\label{sec:conclusions-outlook}
Our proposal on augmenting \acp{bt} with dataflow is, to our knowledge, the
first attempt to do so systematically at the level of a coordination language.
The aim is to combine the best of two worlds that, so far, have seen little
interaction through the involved research communities or in actual practice.
We argue that these concepts can be of mutual benefit.
Compared to ordinary \acp{bt}, our approach improves modularity and ensures
determinism by replacing rather unstructured blackboards with a clean dataflow
notation.
One might argue that this requires some additional modeling effort, but we consider that a matter of sound software engineering that is very likely to pay off, in particular for real-world, complex systems.
Conversely, dataflow formalisms can harness the intuitive, compact \ac{bt} machinery that by now is proven in practice in a large and still growing community of users in game development, robotics control, industrial automation, etc.

With \ac{lf} as the basis for a concrete realization of our proposal, we
leverage its deterministic semantics for concurrent, distributed real-time
systems. Moreover, \ac{lf}'s polyglot nature makes our proposal compatible with
a wide range of target languages.
The combination of \acp{bt} with their simple and intuitive structure and
\ac{lf} with its advanced capabilities in designing complex software systems,
including robust and analyzable timing properties, facilitates the engineering
of reliable software.
Furthermore, we harness \ac{lf}'s pragmatics capabilities, which allow text-first modeling together with automatically generated, customized graphical views.
In particular, a user can choose between the abstract \ac{bt} view, which illustrates the behavioral logic very compactly, and the more detailed dataflow view of the \ac{lf} model, which is synthesized according to the rules presented in \autoref{sec:translation}.
First feedback from our industrial partners at Siemens Technology indicates that
this combination of sound engineering and customizable, automatically
synthesized and always up-to-date graphical views is particularly appealing.

\subsection{Future Work}
As explained in more detail in~\autoref{sec:interface}, a known issue is that target code may use the \ac{bt} incorrectly. In future work, the transformation, detailed in \autoref{sec:translation} could be extended to synthesize additional checks to detect ambiguous output from malformed code and raise an error, or sanitize the output by prioritizing only one of the two states.
As a static solution, the code could be formally verified to adhere to the desired protocol.
Possible language extensions include decorators and memory nodes, as discussed in \autoref{sec:advanced}.
One may also support arbitrary return values in the form of k-\acp{bt}.
Finally, we plan to refine the different graphical views on the model and to further extend the expressiveness of the diagrams.

%
\bibliographystyle{ACM-Reference-Format}
\bibliography{.././bib/cau-rt,.././bib/pub-rts,.././bib/rts-arbeiten,.././local}


\begin{thebibliography}{31}


\ifx \showCODEN    \undefined \def \showCODEN     #1{\unskip}     \fi
\ifx \showDOI      \undefined \def \showDOI       #1{#1}\fi
\ifx \showISBNx    \undefined \def \showISBNx     #1{\unskip}     \fi
\ifx \showISBNxiii \undefined \def \showISBNxiii  #1{\unskip}     \fi
\ifx \showISSN     \undefined \def \showISSN      #1{\unskip}     \fi
\ifx \showLCCN     \undefined \def \showLCCN      #1{\unskip}     \fi
\ifx \shownote     \undefined \def \shownote      #1{#1}          \fi
\ifx \showarticletitle \undefined \def \showarticletitle #1{#1}   \fi
\ifx \showURL      \undefined \def \showURL       {\relax}        \fi
\providecommand\bibfield[2]{#2}
\providecommand\bibinfo[2]{#2}
\providecommand\natexlab[1]{#1}
\providecommand\showeprint[2][]{arXiv:#2}

\bibitem[Agha et~al\mbox{.}(1997)]%
        {AghaMST97}
\bibfield{author}{\bibinfo{person}{Gul Agha}, \bibinfo{person}{Ian~A. Mason},
  \bibinfo{person}{Scott~F. Smith}, {and} \bibinfo{person}{Carolyn Talcott}.}
  \bibinfo{year}{1997}\natexlab{}.
\newblock \showarticletitle{A Foundation for Actor Computation.}
\newblock \bibinfo{journal}{\emph{Journal of Functional Programming}}
  \bibinfo{volume}{7}, \bibinfo{number}{1} (\bibinfo{year}{1997}),
  \bibinfo{pages}{1--72}.
\newblock


\bibitem[Agis et~al\mbox{.}(2020)]%
        {AgisGG20}
\bibfield{author}{\bibinfo{person}{Ramiro Agis}, \bibinfo{person}{Sebastian
  Gottifredi}, {and} \bibinfo{person}{Alejandro Garcia}.}
  \bibinfo{year}{2020}\natexlab{}.
\newblock \showarticletitle{An event-driven behavior trees extension to
  facilitate non-player multi-agent coordination in video games}.
\newblock \bibinfo{journal}{\emph{Expert Systems with Applications}}
  \bibinfo{volume}{155} (\bibinfo{year}{2020}).
\newblock
\urldef\tempurl%
\url{https://doi.org/10.1016/j.eswa.2020.113457}
\showDOI{\tempurl}


\bibitem[Barrera(2018)]%
        {Barrera18}
\bibfield{author}{\bibinfo{person}{Raymundo Barrera}.}
  \bibinfo{year}{2018}\natexlab{}.
\newblock \bibinfo{booktitle}{\emph{Unitiy 2017 {Game} {AI} {Programming} -
  {Third} {Edition}}}.
\newblock \bibinfo{publisher}{Packt Publishing Ltd}.
\newblock


\bibitem[Benveniste et~al\mbox{.}(2003)]%
        {BenvenisteCEH+03}
\bibfield{author}{\bibinfo{person}{Albert Benveniste}, \bibinfo{person}{Paul
  Caspi}, \bibinfo{person}{Stephen~A. Edwards}, \bibinfo{person}{Nicolas
  Halbwachs}, \bibinfo{person}{Paul~Le Guernic}, {and} \bibinfo{person}{Robert
  de Simone}.} \bibinfo{year}{2003}\natexlab{}.
\newblock \showarticletitle{{The Synchronous Languages Twelve Years Later}}. In
  \bibinfo{booktitle}{\emph{{Proc.\ IEEE, Special Issue on Embedded Systems}}},
  Vol.~\bibinfo{volume}{91}. \bibinfo{publisher}{IEEE},
  \bibinfo{address}{Piscataway, NJ, USA}, \bibinfo{pages}{64--83}.
\newblock
\urldef\tempurl%
\url{https://doi.org/10.1109/JPROC.2002.805826}
\showDOI{\tempurl}


\bibitem[Biggar et~al\mbox{.}(2020)]%
        {BiggarZS20}
\bibfield{author}{\bibinfo{person}{Oliver Biggar}, \bibinfo{person}{Mohammad
  Zamani}, {and} \bibinfo{person}{Iman Shames}.}
  \bibinfo{year}{2020}\natexlab{}.
\newblock \showarticletitle{A principled analysis of {Behavior} {Trees} and
  their generalisations}.
\newblock \bibinfo{journal}{\emph{CoRR}}  \bibinfo{volume}{abs/2008.11906}
  (\bibinfo{year}{2020}).
\newblock
\urldef\tempurl%
\url{https://doi.org/10.48550/arXiv.2008.11906}
\showDOI{\tempurl}


\bibitem[Biggar et~al\mbox{.}(2021)]%
        {BiggarZS21}
\bibfield{author}{\bibinfo{person}{Oliver Biggar}, \bibinfo{person}{Mohammad
  Zamani}, {and} \bibinfo{person}{Iman Shames}.}
  \bibinfo{year}{2021}\natexlab{}.
\newblock \showarticletitle{An expressiveness hierarchy of {Behavior} {Trees}
  and related architectures}.
\newblock \bibinfo{journal}{\emph{CoRR}}  \bibinfo{volume}{abs/2104.07919}
  (\bibinfo{year}{2021}).
\newblock
\urldef\tempurl%
\url{https://doi.org/10.48550/arXiv.2104.07919}
\showDOI{\tempurl}


\bibitem[Cola\c{c}o et~al\mbox{.}(2006)]%
        {ColacoHP06}
\bibfield{author}{\bibinfo{person}{Jean-Louis Cola\c{c}o},
  \bibinfo{person}{Gr{\'e}goire Hamon}, {and} \bibinfo{person}{Marc Pouzet}.}
  \bibinfo{year}{2006}\natexlab{}.
\newblock \showarticletitle{Mixing Signals and Modes in Synchronous Data-flow
  Systems}. In \bibinfo{booktitle}{\emph{ACM International Conference on
  Embedded Software (EMSOFT'06)}}. \bibinfo{publisher}{ACM},
  \bibinfo{address}{Seoul, South Korea}, \bibinfo{pages}{73--82}.
\newblock


\bibitem[Cola{\c{c}}o et~al\mbox{.}(2017)]%
        {ColacoPP17}
\bibfield{author}{\bibinfo{person}{Jean{-}Louis Cola{\c{c}}o},
  \bibinfo{person}{Bruno Pagano}, {and} \bibinfo{person}{Marc Pouzet}.}
  \bibinfo{year}{2017}\natexlab{}.
\newblock \showarticletitle{{SCADE} 6: {A} formal language for embedded
  critical software development (invited paper)}. In
  \bibinfo{booktitle}{\emph{11th International Symposium on Theoretical Aspects
  of Software Engineering {TASE}}}. \bibinfo{address}{Sophia Antipolis,
  France}, \bibinfo{pages}{1--11}.
\newblock
\urldef\tempurl%
\url{https://doi.org/10.1109/TASE.2017.8285623}
\showDOI{\tempurl}


\bibitem[Colledanchise and Natale(2018)]%
        {ColledanchiseN18}
\bibfield{author}{\bibinfo{person}{Michele Colledanchise} {and}
  \bibinfo{person}{Lorenzo Natale}.} \bibinfo{year}{2018}\natexlab{}.
\newblock \showarticletitle{Improving the Parallel Execution of Behavior
  Trees}. In \bibinfo{booktitle}{\emph{2018 IEEE/RSJ International Conference
  on Intelligent Robots and Systems (IROS)}}. \bibinfo{pages}{7103--7110}.
\newblock
\urldef\tempurl%
\url{https://doi.org/10.1109/IROS.2018.8593504}
\showDOI{\tempurl}


\bibitem[Colledanchise and Natale(2022)]%
        {ColledanchiseN22}
\bibfield{author}{\bibinfo{person}{Michele Colledanchise} {and}
  \bibinfo{person}{Lorenzo Natale}.} \bibinfo{year}{2022}\natexlab{}.
\newblock \showarticletitle{Handling Concurrency in Behavior Trees}.
\newblock \bibinfo{journal}{\emph{IEEE Transactions on Robotics}}
  \bibinfo{volume}{38}, \bibinfo{number}{4} (\bibinfo{year}{2022}),
  \bibinfo{pages}{2557--2576}.
\newblock
\urldef\tempurl%
\url{https://doi.org/10.1109/TRO.2021.3125863}
\showDOI{\tempurl}


\bibitem[Colledanchise and {\"O}gren(2018)]%
        {ColledanchiseO18}
\bibfield{author}{\bibinfo{person}{Michele Colledanchise} {and}
  \bibinfo{person}{Petter {\"O}gren}.} \bibinfo{year}{2018}\natexlab{}.
\newblock \bibinfo{booktitle}{\emph{Behavior Trees in Robotics and AI: An
  Introduction}}.
\newblock \bibinfo{publisher}{CRC Press}.
\newblock
\showISBNx{9781138593732}
\urldef\tempurl%
\url{https://doi.org/10.1201/9780429489105}
\showDOI{\tempurl}


\bibitem[Conway(1963)]%
        {Conway63}
\bibfield{author}{\bibinfo{person}{Melvin~E. Conway}.}
  \bibinfo{year}{1963}\natexlab{}.
\newblock \showarticletitle{Design of a separable transition-diagram compiler}.
\newblock \bibinfo{journal}{\emph{Commun. ACM}} \bibinfo{volume}{6},
  \bibinfo{number}{7} (\bibinfo{year}{1963}), \bibinfo{pages}{396--408}.
\newblock
\showISSN{0001-0782}
\urldef\tempurl%
\url{https://doi.org/10.1145/366663.366704}
\showDOI{\tempurl}


\bibitem[Dabek et~al\mbox{.}(2002)]%
        {DabekZKMM02}
\bibfield{author}{\bibinfo{person}{Frank Dabek}, \bibinfo{person}{Nickolai
  Zeldovich}, \bibinfo{person}{Frans Kaashoek}, \bibinfo{person}{David
  Mazi\`{e}res}, {and} \bibinfo{person}{Robert Morris}.}
  \bibinfo{year}{2002}\natexlab{}.
\newblock \showarticletitle{Event-Driven Programming for Robust Software}. In
  \bibinfo{booktitle}{\emph{Proceedings of the 10th Workshop on ACM SIGOPS
  European Workshop}} (Saint-Emilion, France) \emph{(\bibinfo{series}{EW 10})}.
  \bibinfo{publisher}{ACM}, \bibinfo{address}{New York, NY, USA},
  \bibinfo{pages}{186--189}.
\newblock
\showISBNx{9781450378062}
\urldef\tempurl%
\url{https://doi.org/10.1145/1133373.1133410}
\showDOI{\tempurl}


\bibitem[Dromey(2003)]%
        {Dromey03}
\bibfield{author}{\bibinfo{person}{R.~Geoff Dromey}.}
  \bibinfo{year}{2003}\natexlab{}.
\newblock \showarticletitle{From Requirements to Design: Formalizing the Key
  Steps}. In \bibinfo{booktitle}{\emph{1st International Conference on Software
  Engineering and Formal Methods ({SEFM} 2003), 22-27 September 2003, Brisbane,
  Australia}}. \bibinfo{publisher}{{IEEE} Computer Society},
  \bibinfo{pages}{2}.
\newblock
\urldef\tempurl%
\url{https://doi.org/10.1109/SEFM.2003.1236202}
\showDOI{\tempurl}


\bibitem[Ghzouli et~al\mbox{.}(2020)]%
        {GhzouliBJD+20}
\bibfield{author}{\bibinfo{person}{Razan Ghzouli}, \bibinfo{person}{Thorsten
  Berger}, \bibinfo{person}{Einar~Broch Johnsen}, \bibinfo{person}{Swaib
  Dragule}, {and} \bibinfo{person}{Andrzej W\k{a}sowski}.}
  \bibinfo{year}{2020}\natexlab{}.
\newblock \showarticletitle{Behavior Trees in Action: A Study of Robotics
  Applications}. In \bibinfo{booktitle}{\emph{Proceedings of the 13th ACM
  SIGPLAN International Conference on Software Language Engineering}} (Virtual,
  USA) \emph{(\bibinfo{series}{SLE 2020})}. \bibinfo{publisher}{Association for
  Computing Machinery}, \bibinfo{address}{New York, NY, USA},
  \bibinfo{pages}{196–209}.
\newblock
\showISBNx{9781450381765}
\urldef\tempurl%
\url{https://doi.org/10.1145/3426425.3426942}
\showDOI{\tempurl}


\bibitem[Grimm et~al\mbox{.}(2020)]%
        {GrimmSSRvH+20}
\bibfield{author}{\bibinfo{person}{Lena Grimm}, \bibinfo{person}{Steven Smyth},
  \bibinfo{person}{Alexander Schulz-Rosengarten}, \bibinfo{person}{Reinhard von
  Hanxleden}, {and} \bibinfo{person}{Marc Pouzet}.}
  \bibinfo{year}{2020}\natexlab{}.
\newblock \showarticletitle{From {Lustre} to Graphical Models and {SCCharts}}.
  In \bibinfo{booktitle}{\emph{Proc.\ Forum on Specification and Design
  Languages (FDL '20)}}. \bibinfo{address}{Kiel, Germany}.
\newblock


\bibitem[Hewitt(1977)]%
        {Hewitt77}
\bibfield{author}{\bibinfo{person}{Carl Hewitt}.}
  \bibinfo{year}{1977}\natexlab{}.
\newblock \showarticletitle{Viewing Control Structures as Patterns of Passing
  Messages}.
\newblock \bibinfo{journal}{\emph{Artif. Intell.}} \bibinfo{volume}{8},
  \bibinfo{number}{3} (\bibinfo{year}{1977}), \bibinfo{pages}{323--364}.
\newblock


\bibitem[Iovino et~al\mbox{.}(2022)]%
        {IovinoSSO+22}
\bibfield{author}{\bibinfo{person}{Matteo Iovino}, \bibinfo{person}{Edvards
  Scukins}, \bibinfo{person}{Jonathan Styrud}, \bibinfo{person}{Petter
  {\"O}gren}, {and} \bibinfo{person}{Christian Smith}.}
  \bibinfo{year}{2022}\natexlab{}.
\newblock \showarticletitle{A survey of Behavior Trees in robotics and AI}.
\newblock \bibinfo{journal}{\emph{Robotics and Autonomous Systems}}
  \bibinfo{volume}{154} (\bibinfo{year}{2022}), \bibinfo{pages}{104096}.
\newblock
\showISSN{0921-8890}
\urldef\tempurl%
\url{https://doi.org/10.1016/j.robot.2022.104096}
\showDOI{\tempurl}


\bibitem[Lohstroh et~al\mbox{.}(2019)]%
        {LohstrohIRGD+19}
\bibfield{author}{\bibinfo{person}{Marten Lohstroh},
  \bibinfo{person}{{\'I}{\~n}igo {\'I}ncer~Romeo}, \bibinfo{person}{Andr\'es
  Goens}, \bibinfo{person}{Patricia Derler}, \bibinfo{person}{Jeronimo
  Castrillon}, \bibinfo{person}{Edward~A. Lee}, {and} \bibinfo{person}{Alberto
  Sangiovanni-Vincentelli}.} \bibinfo{year}{2019}\natexlab{}.
\newblock \showarticletitle{{Reactors: A Deterministic Model for Composable
  Reactive Systems}}. In \bibinfo{booktitle}{\emph{8th International Workshop
  on Model-Based Design of Cyber Physical Systems (CyPhy'19)}},
  Vol.~\bibinfo{volume}{LNCS 11971}. \bibinfo{publisher}{Springer-Verlag},
  \bibinfo{pages}{27}.
\newblock


\bibitem[Lohstroh et~al\mbox{.}(2021)]%
        {LohstrohMBL21}
\bibfield{author}{\bibinfo{person}{Marten Lohstroh}, \bibinfo{person}{Christian
  Menard}, \bibinfo{person}{Soroush Bateni}, {and} \bibinfo{person}{Edward~A.
  Lee}.} \bibinfo{year}{2021}\natexlab{}.
\newblock \showarticletitle{{Toward a Lingua Franca for Deterministic
  Concurrent Systems}}.
\newblock \bibinfo{journal}{\emph{ACM Transactions on Embedded Computing
  Systems (TECS)}} \bibinfo{volume}{20}, \bibinfo{number}{4}
  (\bibinfo{date}{May} \bibinfo{year}{2021}), \bibinfo{pages}{Article 36}.
\newblock
\urldef\tempurl%
\url{https://doi.org/10.1145/3448128}
\showDOI{\tempurl}


\bibitem[Lohstroh et~al\mbox{.}(2020)]%
        {LohstrohMSR+20}
\bibfield{author}{\bibinfo{person}{Marten Lohstroh}, \bibinfo{person}{Christian
  Menard}, \bibinfo{person}{Alexander Schulz-Rosengarten},
  \bibinfo{person}{Matthew Weber}, \bibinfo{person}{Jeronimo Castrillon}, {and}
  \bibinfo{person}{Edward~A. Lee}.} \bibinfo{year}{2020}\natexlab{}.
\newblock \showarticletitle{A Language for Deterministic Coordination Across
  Multiple Timelines}. In \bibinfo{booktitle}{\emph{Proc.\ Forum on
  Specification and Design Languages (FDL '20)}}. \bibinfo{address}{Kiel,
  Germany}.
\newblock
\urldef\tempurl%
\url{https://doi.org/10.1109/FDL50818.2020.9232939}
\showDOI{\tempurl}


\bibitem[Marzinotto et~al\mbox{.}(2014)]%
        {MarzinottoCS14}
\bibfield{author}{\bibinfo{person}{Alejandro Marzinotto},
  \bibinfo{person}{Michele Colledanchise}, \bibinfo{person}{Christian Smith},
  {and} \bibinfo{person}{Peter {\"O}gren}.} \bibinfo{year}{2014}\natexlab{}.
\newblock \showarticletitle{Towards a unified framework for robot control}. In
  \bibinfo{booktitle}{\emph{2014 {IEEE} {International} {Conference} on
  {Robotics} and {Automation} ({ICRA})}}. \bibinfo{pages}{5420--5427}.
\newblock
\urldef\tempurl%
\url{https://doi.org/10.1109/ICRA.2014.6907656}
\showDOI{\tempurl}


\bibitem[Mateas and Stern(2002)]%
        {MaetasS02}
\bibfield{author}{\bibinfo{person}{Michael Mateas} {and}
  \bibinfo{person}{Andrew Stern}.} \bibinfo{year}{2002}\natexlab{}.
\newblock \showarticletitle{A behavior language for story-based believable
  agents}.
\newblock \bibinfo{journal}{\emph{IEEE Intelligent Systems}}
  \bibinfo{volume}{17}, \bibinfo{number}{4} (\bibinfo{year}{2002}),
  \bibinfo{pages}{39--47}.
\newblock
\urldef\tempurl%
\url{https://doi.org/10.1109/MIS.2002.1024751}
\showDOI{\tempurl}


\bibitem[Matsakis and Klock(2014)]%
        {MatsakisK14}
\bibfield{author}{\bibinfo{person}{Nicholas~D. Matsakis} {and}
  \bibinfo{person}{Felix~S. Klock}.} \bibinfo{year}{2014}\natexlab{}.
\newblock \showarticletitle{The rust language}.
\newblock \bibinfo{journal}{\emph{ACM SIGAda Ada Letters}}
  \bibinfo{volume}{34}, \bibinfo{number}{3} (\bibinfo{year}{2014}),
  \bibinfo{pages}{103--104}.
\newblock
\urldef\tempurl%
\url{https://doi.org/10.1145/2692956.2663188}
\showDOI{\tempurl}


\bibitem[Menard et~al\mbox{.}(2023)]%
        {MenardLBC+23}
\bibfield{author}{\bibinfo{person}{Christian Menard}, \bibinfo{person}{Marten
  Lohstroh}, \bibinfo{person}{Soroush Bateni}, \bibinfo{person}{Matthew
  Chorlian}, \bibinfo{person}{Arthur Deng}, \bibinfo{person}{Peter Donovan},
  \bibinfo{person}{Clément Fournier}, \bibinfo{person}{Shaokai Lin},
  \bibinfo{person}{Felix Suchert}, \bibinfo{person}{Tassilo Tanneberger},
  \bibinfo{person}{Hokeun Kim}, \bibinfo{person}{Jeronimo Castrillon}, {and}
  \bibinfo{person}{Edward~A. Lee}.} \bibinfo{year}{2023}\natexlab{}.
\newblock \showarticletitle{High-Performance Deterministic Concurrency using
  Lingua Franca}.
\newblock \bibinfo{journal}{\emph{CoRR}}  \bibinfo{volume}{abs/2301.02444}
  (\bibinfo{date}{January} \bibinfo{year}{2023}).
\newblock
\urldef\tempurl%
\url{https://doi.org/10.48550/arXiv.2301.02444}
\showDOI{\tempurl}
\showeprint[arXiv]{2301.02444}~[cs.PL]


\bibitem[Nilsson(1993)]%
        {Nilsson93}
\bibfield{author}{\bibinfo{person}{Nils Nilsson}.}
  \bibinfo{year}{1993}\natexlab{}.
\newblock \showarticletitle{Teleo-Reactive Programs for Agent Control}.
\newblock \bibinfo{journal}{\emph{Journal of Artificial Intelligence Research}}
   \bibinfo{volume}{1} (\bibinfo{year}{1993}), \bibinfo{pages}{139--158}.
\newblock
\urldef\tempurl%
\url{https://doi.org/10.1613/jair.30}
\showDOI{\tempurl}


\bibitem[Schulz-Rosengarten et~al\mbox{.}(2023)]%
        {SchulzRosengartenvHLB+23}
\bibfield{author}{\bibinfo{person}{Alexander Schulz-Rosengarten},
  \bibinfo{person}{Reinhard von Hanxleden}, \bibinfo{person}{Marten Lohstroh},
  \bibinfo{person}{Soroush Bateni}, {and} \bibinfo{person}{Edward~A. Lee}.}
  \bibinfo{year}{2023}\natexlab{}.
\newblock \showarticletitle{Modal Reactors}.
\newblock \bibinfo{journal}{\emph{CoRR}}  \bibinfo{volume}{abs/2301.09597}
  (\bibinfo{date}{January} \bibinfo{year}{2023}).
\newblock
\urldef\tempurl%
\url{https://doi.org/10.48550/ARXIV.2301.09597}
\showDOI{\tempurl}
\showeprint[arXiv]{2301.09597}


\bibitem[Shoulson et~al\mbox{.}(2011)]%
        {ShoulsonGJM+11}
\bibfield{author}{\bibinfo{person}{Alexander Shoulson},
  \bibinfo{person}{Francisco~M. Garcia}, \bibinfo{person}{Matthew Jones},
  \bibinfo{person}{Robert Mead}, {and} \bibinfo{person}{Norman~I. Badler}.}
  \bibinfo{year}{2011}\natexlab{}.
\newblock \showarticletitle{Parameterizing Behavior Trees}. In
  \bibinfo{booktitle}{\emph{Motion in Games}},
  \bibfield{editor}{\bibinfo{person}{Jan~M. Allbeck} {and}
  \bibinfo{person}{Petros Faloutsos}} (Eds.). \bibinfo{publisher}{Springer
  Berlin Heidelberg}, \bibinfo{pages}{144--155}.
\newblock
\showISBNx{978-3-642-25090-3}
\urldef\tempurl%
\url{https://doi.org/10.1007/978-3-642-25090-3_13}
\showDOI{\tempurl}


\bibitem[Stroustrup(1987)]%
        {Stroustrup87}
\bibfield{author}{\bibinfo{person}{Bjarne Stroustrup}.}
  \bibinfo{year}{1987}\natexlab{}.
\newblock \showarticletitle{What is ``{{Object-Oriented Programming}}''?}. In
  \bibinfo{booktitle}{\emph{ECOOP' 87 European Conference on Object-Oriented
  Programming}}, \bibfield{editor}{\bibinfo{person}{Jean B{\'e}zivin},
  \bibinfo{person}{Jean-Marie Hullot}, \bibinfo{person}{Pierre Cointe}, {and}
  \bibinfo{person}{Henry Lieberman}} (Eds.). \bibinfo{publisher}{Springer
  Berlin Heidelberg}, \bibinfo{address}{Berlin, Heidelberg},
  \bibinfo{pages}{51--70}.
\newblock
\showISBNx{978-3-540-47891-1}


\bibitem[von Hanxleden et~al\mbox{.}(2022a)]%
        {vonHanxledenLF+22}
\bibfield{author}{\bibinfo{person}{Reinhard von Hanxleden},
  \bibinfo{person}{Edward~A. Lee}, \bibinfo{person}{Hauke Fuhrmann},
  \bibinfo{person}{Alexander Schulz{-}Rosengarten},
  \bibinfo{person}{S{\"{o}}ren Domr{\"{o}}s}, \bibinfo{person}{Marten
  Lohstroh}, \bibinfo{person}{Soroush Bateni}, {and} \bibinfo{person}{Christian
  Menard}.} \bibinfo{year}{2022}\natexlab{a}.
\newblock \showarticletitle{Pragmatics twelve years later: a report on {Lingua
  Franca}}. In \bibinfo{booktitle}{\emph{11th International Symposium on
  Leveraging Applications of Formal Methods, Verification and Validation
  (ISoLA)}} \emph{(\bibinfo{series}{Lecture Notes in Computer Science},
  Vol.~\bibinfo{volume}{13702})}. Springer, \bibinfo{address}{Rhodes, Greece},
  \bibinfo{pages}{60--89}.
\newblock
\urldef\tempurl%
\url{https://doi.org/10.1007/978-3-031-19756-7_5}
\showDOI{\tempurl}


\bibitem[von Hanxleden et~al\mbox{.}(2022b)]%
        {vonHanxledenSRA+22}
\bibfield{author}{\bibinfo{person}{Reinhard von Hanxleden},
  \bibinfo{person}{Alexander Schulz-Rosengarten}, \bibinfo{person}{Benjamin
  Asch}, \bibinfo{person}{Soroush Bateni}, \bibinfo{person}{Marten Lohstroh},
  {and} \bibinfo{person}{Edward Lee}.} \bibinfo{year}{2022}\natexlab{b}.
\newblock \bibinfo{title}{A Synchronous View on Behavior Trees}.
\newblock \bibinfo{howpublished}{Presentation at the 29th International Open
  Workshop on Synchronous Programming (SYNCHRON '22), Fr\'ejus, France}.
\newblock
\urldef\tempurl%
\url{https://rtsys.informatik.uni-kiel.de/~biblio/downloads/papers/synchron22.pdf}
\showURL{%
\tempurl}


\end{thebibliography}

\end{document}